\documentclass[twocolumn,prl,aps]{revtex4}

\usepackage[normalem]{ulem}
\usepackage{amssymb}
\usepackage{bm}
\usepackage{epsfig}
\usepackage{graphicx}
\usepackage{amsmath}
\usepackage[latin1]{inputenc}

\def\c60{C$_{60}$}
\def\It{I$_\textrm{t}$}
\def\Tm{T$_\textrm{m}$}
\def\Um{U$_\textrm{m}$}

\def\go{G$_0$}
\def\idec{I$_{\textrm{dec}}$}
\def\Idec{I$_{\textrm{dec}}$}
\def\Vdec{V$_{\textrm{dec}}$}

\def\pdec{P$_{\textrm{dec}}$}
\def\Tdec{T$_{\textrm{dec}}$}
\def\zdec{Z$_{\textrm{dec}}$}
\def\Idec{I$_{\textrm{dec}}$}

\def\Vs{V$_{\textrm{s}}$}

\begin{document}

\title{Resonant electron heating and molecular phonon cooling in single C$_{60}$ junctions}

\author{G. Schulze$^{1}$, K. J. Franke$^{1}$, A. Gagliardi$^{2}$, G. Romano$^{3}$, C. S. Lin$^{2}$, A. Da Rosa$^{2}$, T. A. Niehaus$^{2}$, Th. Frauenheim$^{2}$, A. Di Carlo$^{3}$, A. Pecchia$^{3}$, J.I. Pascual$^{1}$}

\affiliation{$^1$ Inst. f\"ur Experimentalphysik, Freie
Universit\"at Berlin, Arnimallee 14, 14195 Berlin, Germany \\
$^2$ Bremen Center for Computational Materials Science, University
of Bremen, D-28359, Germany \\
$^3$ Università di Roma 'Tor Vergata', 00133 Roma, Italy }

\begin{abstract}

We study heating and heat dissipation of a single C$_{60}$ molecule
in the junction of a scanning tunneling microscope (STM) by
measuring the electron current required to thermally decompose the
fullerene cage. The power for decomposition varies with electron
energy and reflects the molecular resonance structure. When the STM
tip contacts the fullerene the molecule can sustain much larger
currents. Transport simulations explain these effects by molecular
heating due to resonant electron-phonon coupling and molecular
cooling by vibrational decay into the tip upon contact formation.

\end{abstract}

\date{\today}

\maketitle

The paradigm of molecular electronics is the use of a single
molecule as an electronic device \cite{Aviram74}. This concept is
sustained on the basis that a single molecule (or a molecular thin
film) should withstand the flow of electron current densities as
large as 10$^{10}$ A/m$^2$ without degrading. A fraction of these
electrons heat the molecular junction through inelastic scattering
with the molecule \cite{NitzanJPCM07}. The temperature at the
junction is a consequence of an equilibrium between heating due to
electron flow and heat dissipation out of the junction. The former
is dominated by the coupling of electronic molecular states with
molecular vibrons \cite{NitzanJPCM07,NitzanPRB07,PecchiaPRB07}. The
latter depends on the strength of the vibrational coupling between
the ``hot" molecular vibrons and the bath degrees of freedom of the
"cold" electrodes.

Theoretical studies predicted that current-induced
heating in molecular junctions can be large enough to affect the
reliability  of molecular devices \cite{NitzanJPCM07}. However,
experimental access to this information is very limited. Recent
studies of the thermally activated force during molecular detachment
from a lead \cite{HuangNL06,HuangNN07} and of structural fluctuation
during attachment to it \cite{BerndtPRL07} reveal that  the
temperature of a molecular junction can reach several hundred
degrees under normal working  conditions, thus revealing that
present devices work on the limit of practical operability
\cite{Tour00}. Heat dissipation away from the junction becomes an important issue.

In this work, we characterize the mechanisms of heating and heat dissipation
induced by the flow of current across a single molecule. Our approach is based
on detecting the limiting electron current inducing molecular
decomposition at varying applied source-drain bias (i.e. the maximum power one molecule can sustain). We use a low
temperature scanning tunneling microscope (STM) to control the flow
of electrons through a single C$_{60}$ molecule at an increasing
rate until the molecule decomposes. By comparing the power applied
for decomposition (\pdec) in tunneling regime and in contact with the
STM tip we find that it depends significantly on two factors: i)
\pdec\ decreases when molecular resonances
participate in the transport, evidencing that they enhance the
heating; ii) \pdec\ increases as the molecule is contacted to the source and drain electrodes, revealing the heat dissipation by phonon coupling to the leads.
A good contact between the single-molecule (SM)
device and the leads is hence an important requirement for its
operation with large current densities.

\begin{figure} [tb]
\begin{center}
\includegraphics[width=.95\columnwidth]{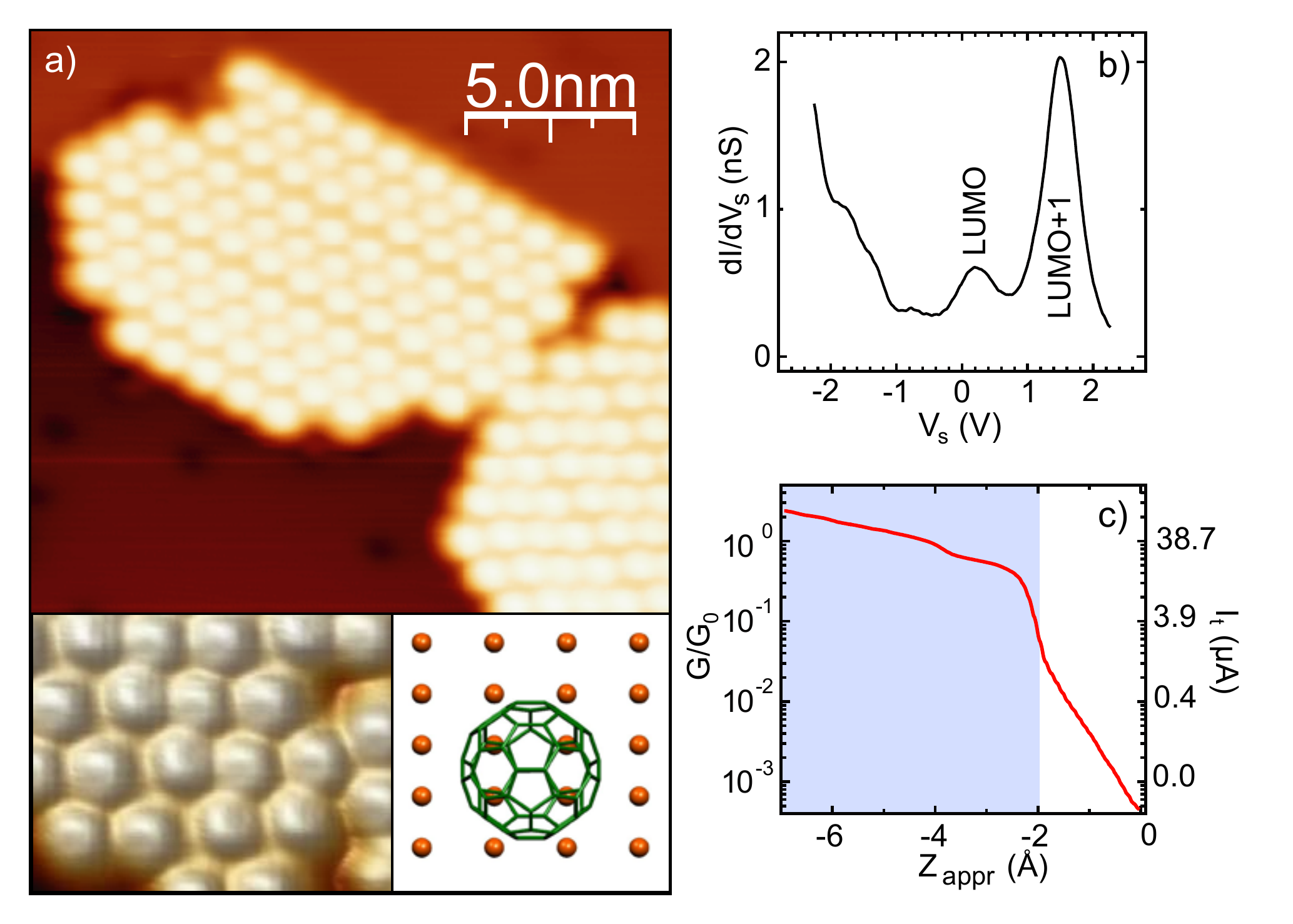}
\end{center}
\caption{(a) STM image of a 0.2 monolayer film of C$_{60}$ on
Cu(110)   (\It=1.1 nA, \Vs=1.75 V). The intramolecular structure
(inset; \It=1.0 nA, \Vs=2.25 V, \cite{wsxm}) reveals an intrinsic
asymmetry consistent with the adsorption orientation from DFT
simulations (right inset). (b) Differential conductivity spectrum of
a C$_{60}$\ molecule measured using a lock-in amplifier
(R$_{\textrm{junct}}$=2.2 G$\Omega$, V$_{\textrm{ac}}$=20 mV rms).
The structure in the spectrum is due to  C$_{60}$\ electronic
resonances. (c) Conductance and current vs. Z$_{\textrm{appr}}$ plot
on top of a C$_{60}$\ molecule (\Vs=0.5 V). The shaded area
indicates contact regime.} \label{STMim}
\end{figure}

Our experiments are carried out in  a custom-made ultra-high vacuum
STM at a temperature of 5 K. We choose a Cu(110) single crystal
surface  because here C$_{60}$  adsorbs in a well-defined
configuration, between 4 of the top most Cu atoms \cite{FaselPRB99}.
By annealing a sub-monolayer film of C$_{60}$ to 470 K we produce
ordered fullerene islands with a pseudo-hexagonal structure (Fig.
\ref{STMim}(a)), in which C$_{60}$ adsorbs keeping a
pentagon-hexagon C-C bond pointing upwards. Scanning tunneling
spectroscopy (STS) (Fig. \ref{STMim}(b)) shows a clear spectroscopic
fingerprint characterized by a strong resonance at 1.5 eV above the
Fermi level, and associated to the alignment of the LUMO+1 resonance
(LUMO: lowest unoccupied molecular orbital). The LUMO resonance
appears as a weaker broader peak around 0.2 eV and is partially
occupied \cite{note2}. A spectrum like the one shown in Fig.
\ref{STMim}(b) is thus taken here as a litmus test for the integrity
of the C$_{60}$ molecule.

In our experiment, we approach the STM tip \cite{notetip} a distance
Z$_{\textrm{appr}}$ towards a single C$_{60}$ at constant sample
bias (\Vs) and record the current thus flowing through the  molecule
(I(Z$_{\textrm{appr}}$)). The tunnel regime (identified here as the
exponential regime in the I(Z$_{\textrm{appr}}$) plots) extends
until the junction conductivity reaches $\sim$ 0.03 \go\ (\go =77.5
$\mu$S). Beyond this point, a tip-C$_{60}$\ contact starts to be
formed, and the I(Z$_{\textrm{appr}}$) plots deviate smoothly from
the exponential behavior \cite{GimzPRL95,BerndtPRL07}. C$_{60}$ is
very stable under the proximity of the STM tip. If a small positive
sample bias (\Vs$<$0.6) is used, C$_{60}$ withstands tip contact and
indentations of several \AA ngstroms, holding currents close to
100$\mu$A (Fig. \ref{STMim}(c)).

\begin{figure} [tb]
\begin{center}
\includegraphics[width=.9\columnwidth]{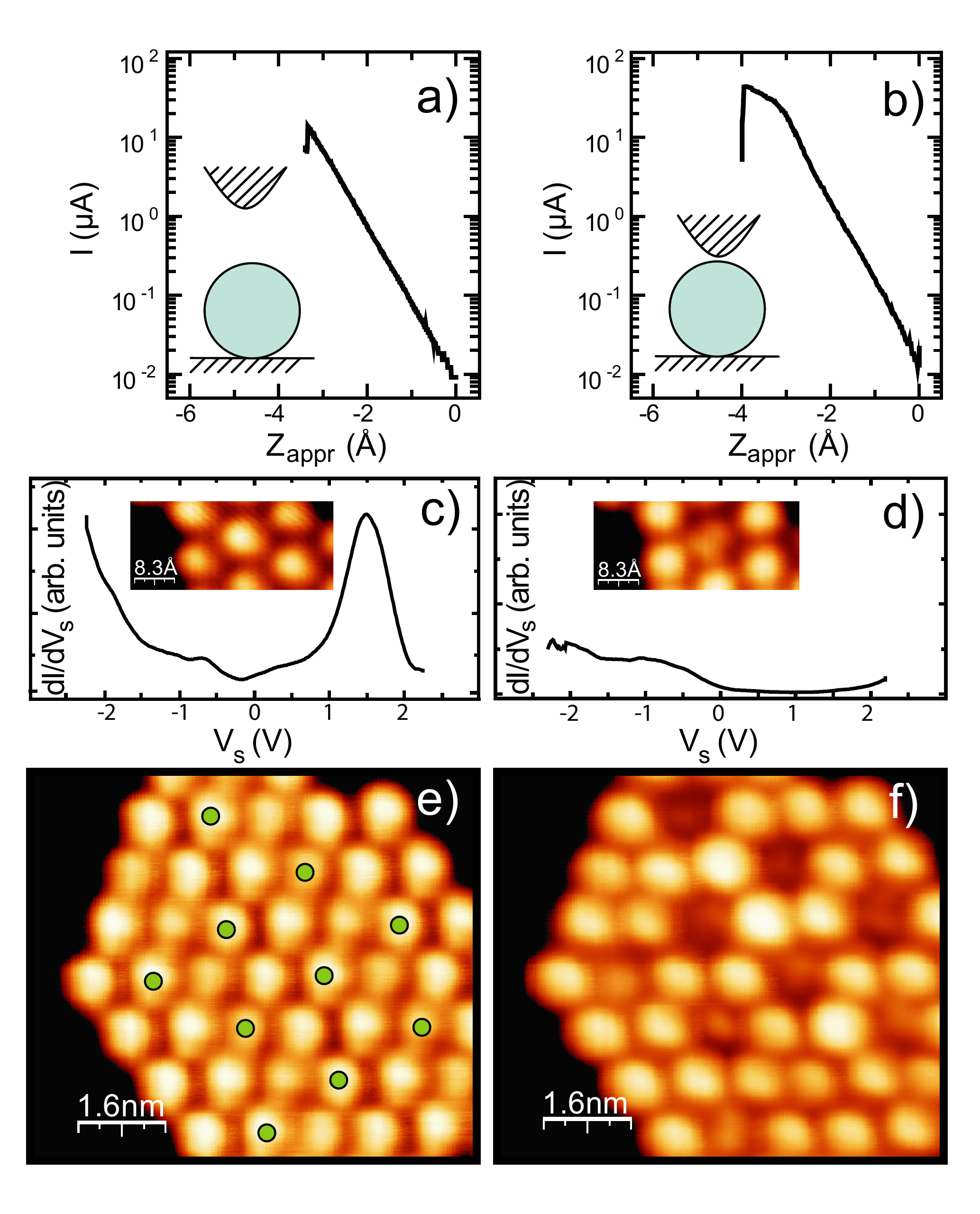}
\end{center}
\caption{(Color online)   I(Z$_{\textrm{appr}}$) plots of C$_{60}$\
showing molecular decomposition in (a) tunnel regime (\Vs=2.0 V) and
(b) after forming a tip-C$_{60}$\ contact (\Vs=1.0 V, acquisition
time t=1.5s). (c) STS spectrum of a molecule before  and (d) after
an I(Z$_{\textrm{appr}}$) plot like in panel (a). The resonance
structure disappears, thus indicating destruction of the fullerene
cage. (e) Island of C$_{60}$\ molecules before and (f) after
performing several I(Z$_{\textrm{appr}}$) events on the molecules
marked in (e). After this, all the marked molecules show a lower
height and a dI/dV spectrum similar to panel (d) (\It=1.0 nA,
\Vs=2.25 V). } \label{inden}
\end{figure}

When the value of \Vs\ is increased above 0.6 V, C$_{60}$\ degrades
during the tip approach, as we could identify from three facts (Fig.
\ref{inden}): i) a sharp discontinuity in the I(Z$_{\textrm{appr}}$)
plots indicates an irreversible change in the molecule;  ii) the
height of the degraded molecule is typically more than 1 \AA\ lower
than its neighbors; iii) the dI/d\Vs\ spectrum reveals that the
characteristic resonance structure vanishes. The precise way in
which C$_{60}$\ is degraded can not be determined exactly in our
measurements. The disappearance of molecular resonances in the
spectra hints that the most probable result is a rupture of the
icosahedral carbon cavity. The decomposition is observed solely on
the molecule selected for the indentation. Thus, we can discard that
an electron-induced polymerization with neighboring molecules takes
place \cite{ZhaoAPL94}.

\begin{figure} [tb]
\begin{center}
\includegraphics[width=1.0\columnwidth]{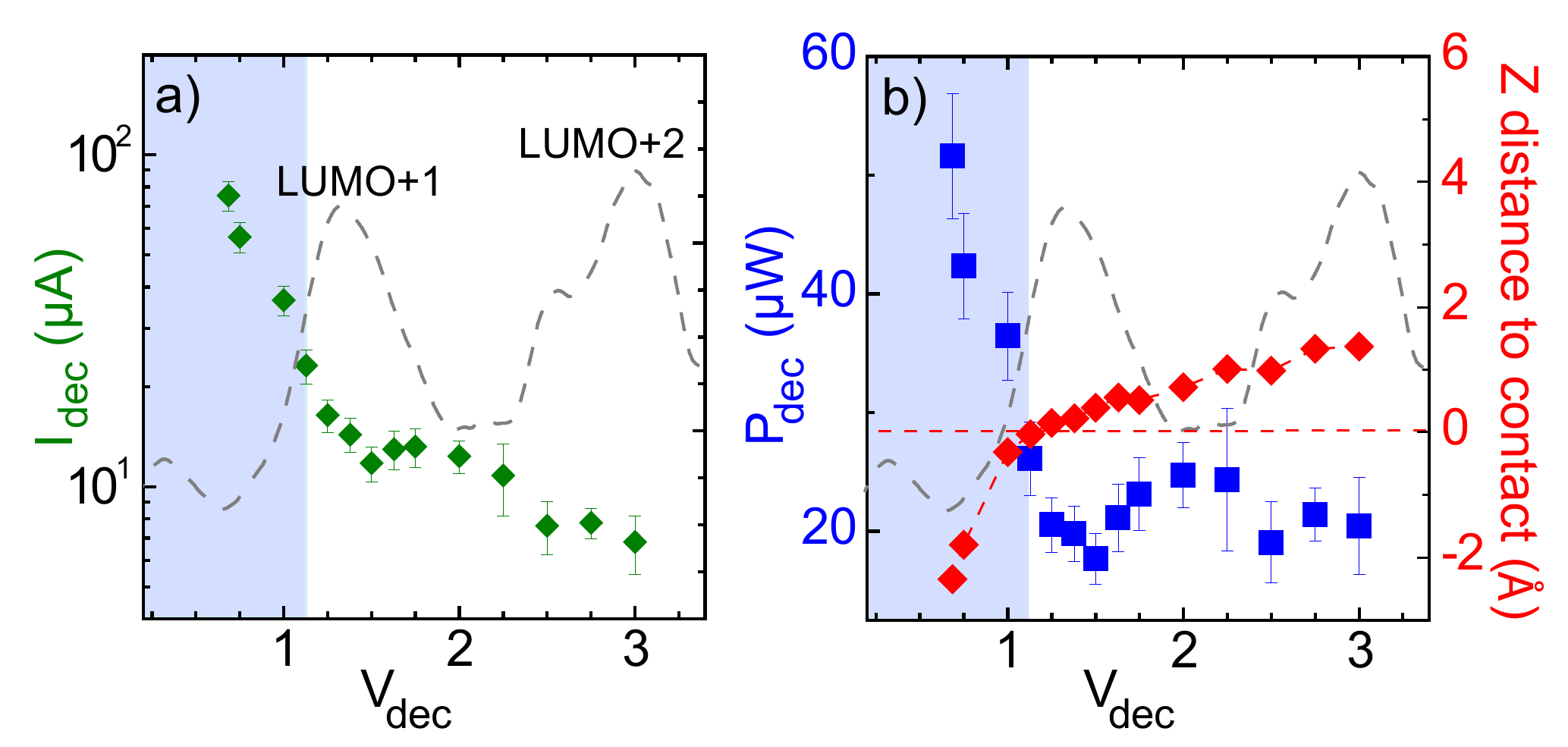}
\end{center}
\caption{(Color online) (a) Statistical average of current reached
at the point of decomposition \idec\ vs. sample bias \Vs. Two sigma
error bars are indicated.  All the events considered in the
statistical average (180) correspond to degradation as in Fig. 2,
and were done on fullerenes surrounded by unperturbed C$_{60}$\
molecules. (b) Bias dependence of \pdec\ (squares, left scale) and
distance to contact \zdec\ (diamonds, right scale). For events in
tunnel regime, \zdec\ is obtained by a linear extrapolation. Shaded
areas indicate decomposition in contact regime. Gray dashed lines
represent normalized conductance (dI/d\Vs) plots, indicating the
resonance alignment.} \label{stat}
\end{figure}

A current drop as in  Figs. \ref{inden}(a) and \ref{inden}(b)
provides the position \zdec\ and current \idec\ at which molecular
decomposition occurs. Both  \zdec\  and  \idec\ depend strongly on
the bias value \Vs\ used. In Fig. \ref{stat}(a) we plot the
statistical average of \idec\ as a function of \Vs\ between 0.6 V to
3.0 V. For \Vs= 3.0 V an electron current \idec = 7 $\mu$A suffices
to produce the C$_{60}$\ degradation. This occurs when the tip is
still more than 1 {\AA} away from the contact position, i.e. in
tunnel regime. As \Vs\ is reduced \idec\ increases gradually (with a
small plateau around 1.5 eV and 2.5 eV) and, accordingly
decomposition occurs with the tip closer to the fullerene, but still
in the tunnel regime (i.e. with I(Z$_{\textrm{appr}}$) curves as in
Fig. \ref{inden}(a)). For \Vs\ below $\sim$1.2 V a more pronounced
increase of \idec\ is observed when approaching the tip. In this
range the onset of tip-C$_{60}$\ contact is already detected in the
I(Z$_{\textrm{appr}}$) plots (as in Fig. \ref{inden}(b)). Electron
currents as high as 70 $\mu$A can flow through the molecular
junction for the lowest bias achieving degradation (\Vs $\sim$ 0.6
V). In this regime the STM tip indents more than 2 \AA\ beyond the
contact position. However, we can rule out a mechanical process of
rupture because below the threshold bias of 0.6 V C$_{60}$\ remains
unaffected upon tip indentations of more than 4 {\AA} (see Fig.
\ref{STMim}(c)). Instead, we note that C$_{60}$\ undergoes a thermal
decomposition on metal surfaces at temperatures around 1000 K
\cite{CepekPRB96, PascualSS97,Swami99, Saltas01}. Thus, we consider
that the decomposition of C$_{60}$\ is a \textit{current-driven}
thermal process, where the critical current depends crucially on the
applied voltage.

The two different regimes of molecular decomposition in tunnel and
in contact as suggested by the behavior of \idec\ with \Vs\ in Fig.
\ref{stat}(a) become more evident when we plot the power \pdec\
(\pdec $=$ \idec $\times$ \Vs) applied to the C$_{60}$\ junction for
its degradation (Fig. \ref{stat}(b)). In tunnel regime, i.e. above
\Vs\ $\sim$ 1.2 V, \pdec\ amounts to $\sim$ 20 $\mu$W and shows
small oscillations at the LUMO+1 and LUMO+2 position. For lower bias
\pdec\ increases sharply up to more than 50 $\mu$W, reflecting that
here the junction can sustain larger current densities. It should be
noted that in this regime decomposition is attained when the tip is
in contact with the molecule. Hence, the very sharp change in \pdec\
as compared to the oscillations in \pdec\ around higher molecular
resonances suggests that the contact plays an important role for the
decomposition process.

The origin of molecular dissociation in the two different transport
regimes can be rationalized from a conceptual definition of
molecular temperature. The electron tunneling rates used to
decompose the C$_{60}$\ molecule (a few fs$^{-1}$) are larger than
typical phonon decay rates for adsorbate systems. Hence, we consider
a current-induced vibrational heating, in which C$_{60}$\ vibrations
are excited in a non-equilibrium distribution \cite{Salam94, Gao97}.
For a certain set of current and bias values, the total vibrational
energy \Um\ stored in the molecule (and hence its temperature \Tm)
depends on the balance between the \textit{heat generated} by the
inelastic scattering of electrons with molecular modes, and
\textit{heat dissipation} into the ``cold" leads (T=5K)
\cite{PecchiaPRB07,NitzanJPCM07, PecchiaJCM07}. Following this
picture, a larger degradation power \pdec\ is due to either a less
effective heat generation or to a more effective dissipation of heat
into the leads.

In order to analyze and corroborate the influence of the resonances
and the tip-molecule contact theoretically we have performed model
transport calculations based on the Non-Equilibrium Green's Function
(NEGF) formalism \cite{Pecchia2, Gagliardi}. The C$_{60}$ molecule
is relaxed on a slab of 8 layers of copper in a 4 x 5 unit cell
using density functional theory calculations \cite{vdWaals}. The
resulting structure is in good agreement with experiments
\cite{FaselPRB99}. Unoccupied molecular states (LUMO, LUMO+1 and
LUMO+2) are located at 0.7 V, 1.5 V and 3.0 V above the Fermi level,
respectively. The tip is represented by a Cu atom adsorbed on a
jellium surface.

Inelastic electron scattering in C$_{60}$ is calculated using the
Self Consistent Born Approximation (SCBA) \cite{BAapprox}. The
non-equilibrium phonon population, $\textrm{N}_\textrm{{q}}$ (where
the q index runs over all 174 normal modes of vibration of the
C$_{60}$), is deduced from a rate equation including the phonon
absorption (A$_{\textrm{q}}$) and emission probabilities
(E$_\textrm{q}$) in the device calculated as in ref.
\cite{PecchiaJCM07}. A ``model" molecular temperature, \Tm, can be
associated to the internal energy,
$\textrm{U}_{\textrm{m}}=\sum_{\textrm{q}} h
\nu_{\textrm{q}}\textrm{N}_\textrm{{q}}$, by assuming a
Bose-Einstein phonon population that produces the same internal
energy as the non-equilibrium population $\textrm{N}_\textrm{{q}}$
\cite{note1}.

\begin{figure} [tb]
\begin{center}
\includegraphics[width=.95\columnwidth]{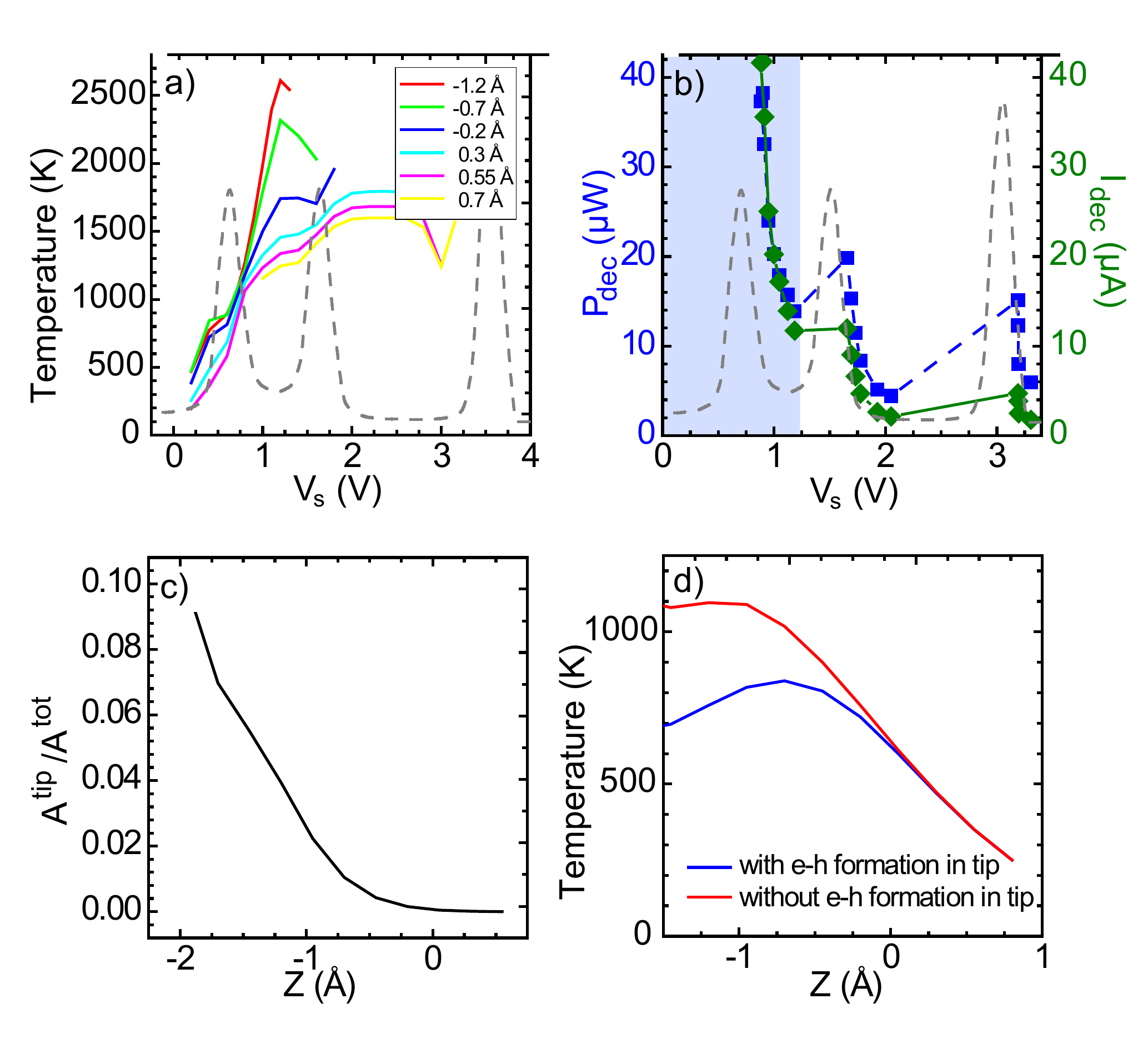}
\end{center}
\caption{(color online) Results of theoretical simulations: (a)
Temperature \Tm\ of the molecule vs. \Vs\ for the indicated
distances to contact.  The oscillations of \Tm\ are associated to
the molecular resonances (dashed line). (b) The values \Idec
(diamonds, green continuous line) and \pdec (squares, blue broken
line) are marked for a threshold temperature of 1650 K of the
C$_{60}$ molecule. The shaded area marks the contact regime. (c)
Ratio of phonon absorption rate (A$^{\textrm{tip}}$) to total
absorption rate in the junction (A$^{\textrm{tot}}$) vs.
tip-molecule distance (tip-molecule contact is defined as 0 \AA) at
0.4 V.  (d) Temperature \Tm\ at 0.4 V vs. tip-molecule distance with
(blue) and without (red) electron-hole pair formation in the tip.
\label{theo}}
\end{figure}

Fig. 4 (a) shows the effect of increasing the applied bias on the
molecular temperature \Tm. For the experimental values of
tip-molecule distance, the applied bias leads to a heating of the
molecule up to temperatures above 1000 K. Such  high temperatures
are reached with only a small fraction of electrons being
inelastically scattered by  molecular vibrons  ($\sim$ 10$^{-3}$).
The heating becomes more effective when the LUMO resonance enters in
conduction due to resonant electron-phonon emission
\cite{PecchiaPRB07}. The temperature increase is largest when all
C$_{60}$ vibrational modes can be excited, which lie in a band of
200 meV width above the resonance. A similar rise in \Tm\ appears at
bias values when higher order resonances enter into conduction.
Contrary to this, phonon assisted tunneling contributes to cooling
the C$_{60}$ molecule right below the resonances, causing in total a
modulated rise in \Tm.

To simulate the experimental curves of Fig. 3, we set a critical temperature for decomposition, \Tdec, and obtain from Fig. \ref{theo}(a) the corresponding set of values \Vdec\ and \zdec. From the I-V characteristics we then extract the corresponding \Idec. For \Tdec\ $\sim$ 1650 K the applied power is similar to the experimental values \cite{note5}. For bias values larger than 1.2 V the decomposition takes place in the tunnel regime and a step-like behavior is obtained for \Idec. Where \Idec\ remains constant \pdec\ increases, thus producing the oscillations shown in Fig. \ref{theo}(b). These reflect the resonant phonon cooling and heating below and above a resonance, respectively. Such oscillations resemble the experimental oscillations  of \pdec\ at the LUMO+1 and LUMO+2, hence confirming
a resonance mediated mechanism of molecular heating in tunneling.

This, however, does not explain the very large increase in \pdec\
for sample bias below 1.2 V. Here, as in the experiment, molecular
decomposition is achieved once the tip is in contact with C$_{60}$.
The formation of a tip-molecule contact enhances the dissipation of
vibrations from the hot molecule into the cold tip. There are two
possible mechanisms of mode quenching: vibrational decay due to
vibron-phonon coupling and to electron-hole (e-h) pair excitations.
Our calculations also suggest that the former plays a small role in
the cooling effect, since the phonon bands at the leads are much
narrower than the 200 meV wide C$_{60}$\ vibrational spectrum. On
metal surfaces, instead, the main mechanism is the e-h excitation in
the leads \cite{Gao92}. From our calculations we can estimate the
contribution of e-h pair creation in the tip by extracting the
corresponding phonon absorption rate (A$_\textrm{q}^{\textrm{tip}}$)
\cite{noteAq}. Molecular mode quenching becomes only important in
the proximity of contact formation and further indentation in the
molecule as revealed by the monotonous increase of
A$_\textrm{q}^{\textrm{tip}}$ (Fig.4(c)). The effect on \Tm\ can be
determined by setting A$_\textrm{q}^{\textrm{tip}}$ to zero and
calculating again the effective temperature. Fig. \ref{theo}(d)
evidences that the temperature thus determined increases much faster
when approaching the tip towards the molecule than in the case where
e-h excitation in the tip is allowed. Hence, our calculations
strongly support that the sharp increase in both \Idec\ and \pdec\
result from phonon cooling upon contact formation. At low bias, the
LUMO resonance starts to be removed from the conduction window and,
as shown in Fig. \ref{theo}(a), the molecular temperature decreases
drastically. This explains that the thermal decomposition cannot be
achieved in the experiment below 0.6 eV.

In summary, our results reveal that resonant tunneling through
molecular states in a single molecule device can generate sufficient
heat to thermally decompose the molecular junction. For the case of
a C$_{60}$ molecule on a Cu(110) surface, a power of only 20 $\mu$W
is sufficient. In order to increase the current density a molecular
junction can sustain, it would be useful to remove molecular
resonances from the transport window. However, in most cases the
molecular device properties rely crucially on molecular resonances,
and hence cannot be engineered without loosing the functionality.
Good contact of the molecule with the leads then opens the
possibility for the single molecule device to withstand larger
current densities.

This research was supported by the Deutsche Forschungsgemeinschaft,
through the collaborative projects SPP 1243 and SFB 658.


\begin{thebibliography}{99}
\bibitem{note1} We approximate the molecular by fitting the total atomic kinetic energy to a Bose-Einstein
distribution at temperature \Tm. A better approach employs a fictitious
external phonon bath to "measure" \Tm\ \cite{NitzanJPCM07}.
Qualitatively, both methods are similar.
\bibitem{note5} No qualitative difference is seen in the results if \Tdec\ is increased by several hundreds of degrees or by using U$_\textrm{m}$ as a critical parameter for molecular decomposition.
\bibitem{noteAq} The excitation of an e-h pair in the tip can be seen as a real process
in which an electron from the tip absorbs a phonon and is reflected
back into the tip:
A$_q^{tip} = 2/h \int_{-\infty}^\infty Tr[R_t \times \alpha_q \times R_t \times \alpha_q][f_t(E-\hbar w_q) (1 - f_t(E))]dE $,
where $\alpha_q$ is the electron-phonon coupling matrix, $R_t$ is the projected density of states on C$_{60}$+tip states, and $f_t$ is the tip Fermi function.
\end{thebibliography}
\end{document}